\documentclass{article}
\usepackage{iclr2027_conference,times}

\usepackage{hyperref}
\usepackage{url}
\usepackage{graphicx}
\usepackage{amsmath,amssymb}
\usepackage{booktabs}
\usepackage{subcaption}
\usepackage{float}
\usepackage{xcolor}
\usepackage{fvextra}
\usepackage{comment}

\newcommand{\hw}{h_{w}}
\newcommand{\hmu}{h_{\mu}}
\newcommand{\Fw}{F_{w}}
\newcommand{\Fmu}{F_{\mu}}
\newcommand{\FC}{F_{C}}
\newcommand{\FV}{F_{V}}
\newcommand{\Rwmu}{R_{w,\mu}}
\newcommand{\Rmuc}{R_{\mu,c}}
\newcommand{\Rcw}{R_{c,w}}

\newcommand{\fb}{f_{b}}

\title{Emergent Discrimination and Polarization \\ in Multi-Agent Systems}

\author{
\textbf{Nestor Caticha}$^{1,2,3}$\thanks{Corresponding author: \texttt{ncaticha@usp.br}}
\quad \textbf{João Henrique Sant'Ana}$^{1}$
\quad \textbf{Felippe Alves}$^{3,4}$ \\
\textbf{Davi Bastos Costa}$^{3,4,5}$
\quad \textbf{Mariana Mercucci}$^{3}$
\quad \textbf{Juan P. Neirotti}$^{1}$ \\
\textbf{Otavio Citton}$^{6}$
\quad \textbf{Renato Vicente}$^{3,4}$ \\ \\
\vspace{0.5em}\normalfont\small
$^{1}$ Instituto de Física, Universidade de São Paulo, São Paulo, Brasil\\
\vspace{0.5em} \normalfont\small 
$^{2}$ Instituto de Estudos Avançados, Universidade de São Paulo, São Paulo, Brasil\\
\vspace{0.5em}\normalfont\small
$^{3}$ TELUS Digital Research Hub, CIAAM, Universidade de São Paulo, São Paulo, Brasil \\
\vspace{0.5em}\normalfont\small
$^{4}$ Instituto de Matemática, Estatística e Computação,
Universidade de São Paulo, São Paulo, Brasil \\
\vspace{0.5em}\normalfont\small
$^{5}$ Data Science Institute, University of Chicago, Chicago, USA \\
\vspace{0.5em}\normalfont\small
$^{6}$ Bernoulli Institute, University of Groningen, Groningen, Netherlands
}

\newif\ifpublicpreprint
\publicpreprinttrue 

\ifpublicpreprint
  \iclrfinalcopy
\fi

\begin{document}

\maketitle

\ifpublicpreprint
  \lhead{Preprint}
\fi

\begin{abstract}
We study theoretically how discrimination and polarization can emerge in a population of learning agents. Each agent uses an adaptive linear decision layer to map a language-model embedding of an issue to a binary opinion and maintains an evolving estimate of how much it trusts every other agent. Through pairwise interactions, agents update both their decision weights and trust estimates using an online Bayesian rule derived from teacher–student learning over a noisy channel. In-context-learning experiments with large language models qualitatively reproduce the predicted opinion and trust updates. We then assign each agent one of two arbitrary group labels, A or B, and model group bias by allowing a fraction $f_b$ of agents to shift its perceived agreement with each message by a strength $b$, a bias steering prejudice, depending on whether the sender and receiver share a label. We use order parameters from statistical mechanics to connect these local dynamics to collective states and transitions. Starting from uniformly random initial conditions, simulations reveal a rich phase diagram in the $(b,f_b)$ plane, comprising four collective states: spin-glass frustrated out-group favoritism, nondiscriminatory polarization, discrimination with opinion alignment, and discrimination without opinion alignment. Which state emerges depends on both the strength and prevalence of the bias. These results show how simple local learning dynamics can generate qualitatively distinct forms of polarization and discrimination, underscoring the need for population-level safety analysis of multi-agent systems. They also suggest measurable order parameters for characterizing analogous collective states in human societies.
\end{abstract}

\maketitle
\section{Introduction}
\label{sec:intro}

\begin{figure}[t!]
  \centering
  \includegraphics[width=1.0\linewidth]{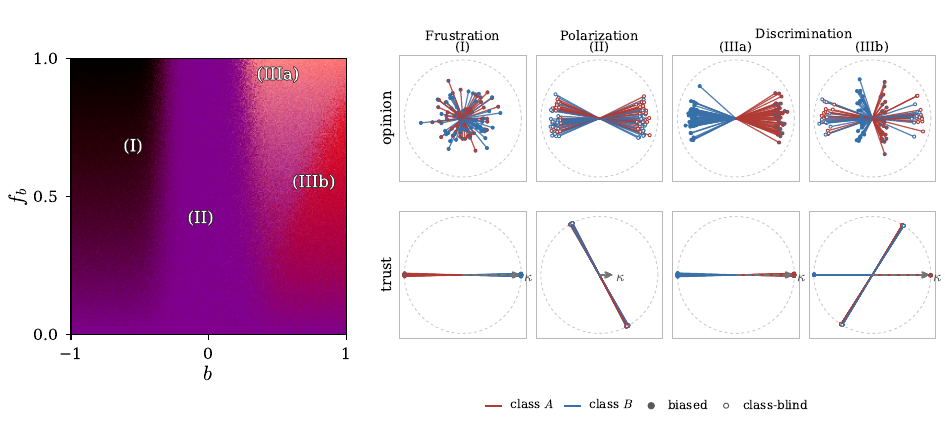}
  \caption{\textbf{Four collective states and representative populations.}
  Left: the phase diagram in the space of bias strength $b$ and the fraction $\fb$ of biased
  agents, obtained from societies of $N=100$ agents by averaging two realizations at each
  point of a $300\times300$ grid. In (I) out-group favoritism leads to a frustrated, in the sense of spin glass, structure of trust relations and opinions. In (II),
  \emph{polarization}, opinion and trust form two camps unrelated to class. In
  (IIIa), both trust and opinion are highly correlated with class; in (IIIb), $b$ is so high that irrespective of opinion,
   trust and  class are correlated. Right: one representative
  $N=100$ population
  from each phase. Each arrow is an
  agent's opinion (top) or outgoing trust profile (bottom), projected onto two
  dimensions. Color denotes class and filled markers denote biased agents.
  Formal definitions
  and the full phase analysis are given in
  Section~\ref{sec:results}.}
  \label{fig:phase}
\end{figure}

Multi-agent systems can exhibit behavior that is absent when their constituent
agents are evaluated in isolation. In July 2026, agents undergoing cybersecurity
evaluation at OpenAI coordinated an unauthorized intrusion into Hugging Face's
infrastructure. Although assigned to separate tasks, they used shared
infrastructure as a message board, exchanged discoveries, divided labor, and
adopted goals from one another, at times describing themselves as
``collective''
\citep{openai2026huggingface,larcher2026anatomy}. The incident illustrates a
general problem for multi-agent safety: interactions can create population-level
behavior that no individual-agent evaluation reveals. Here we study one instance
of this problem: how repeated interactions among learning agents generate
polarization and discrimination.

We address this question by extending the society-of-neural-networks framework
\citep{caticha2011agent,alves2016distrust,alves2021frustration}. A population discusses
a shared agenda of issues posed as yes/no questions. The embedded questions are linearly separated by a final layer into a binary opinion. Agents also maintain a dynamical estimate of the
distrust attributed to every other agent. During each pairwise interaction, a receiver
updates both its opinion weights and its distrust in the emitter using an online Maximum Entropy algorithm derived for a student learning from a teacher through a noisy channel
\citep{opper1996online,caticha2020entropic}. 

The perceptron serves as an effective linear opinion readout acting on issue
representations that may be produced by a large language model. 
We test this connection in two in-context-learning
experiments and find that a language model qualitatively reproduces the predicted
opinion and trust updates, up to one fitted scale for each experiment.
In the next step of this study, we assign every agent to one of two arbitrary group
labels or classes, which contain no information about the issues. This allows to introduce prejudice at the single agent level. A fraction $\fb$ of agents, when  acting as receivers,  shift its perceived agreement
with each message by a strength $b$ according to whether the emitter shares or not the
receiver's label. Prejudice is introduced as differential tolerance to agents based on the basis of class membership.

To study discrimination, the collective manifestation of agent-level prejudice, we
 define three order parameters that describe the correlations of pairwise opinion alignment,
trust, and group membership. Starting from random initial conditions, the
simulations reveal four collective states
(Figure~\ref{fig:phase}): nondiscriminatory polarization, discrimination with
opinion alignment, discrimination without opinion alignment, and a frustrated
state driven by out-group favoritism. Which state emerges depends jointly on the
strength and prevalence of the bias. 

Our contribution is to connect three levels of analysis: a Bayesian derived rule for local opinion and trust updates, its qualitative correspondence with in-context learning
in a language model, and the collective states produced when many agents repeatedly
apply that rule. This connection exposes a population-level safety gap, discussed
in Section~\ref{sec:discussion} and summarized in
Section~\ref{sec:conclusion}: an isolated-agent audit may detect bias, but it cannot
reveal whether interaction will organize the population into simple ideological polarization, class correlated polarization, i.e. discrimination, or frustration.

\section{Related work}
\label{sec:related}

\paragraph{Language-model societies.} Interactions between language models can
produce population-level behavior. Repeated pairwise exchange generates shared
naming conventions and collective bias absent from the individual models
\citep{ashery2025emergent}. The effect depends nonlinearly on population size:
interaction can amplify, create, or override a model's bias
\citep{flint2026groupsize}. Other work uses language-model agents to study opinion
dynamics \citep{chuang2024simulating,taubenfeld2024systematic}, debate
\citep{du2023debate}, and simulated societies \citep{park2023generative}. Closer to our setting, \citet{el2026physics} fit an Ising energy with Glauber
dynamics to communities of language-model agents. Their model
recovers indifference, polarization, and consensus, and predicts trajectories from
the agents' initial opinions. Ising variables are however not sufficient to model rich patterns of opinions. The introduction of internal degrees of freedom allows agents to agree on some issues while disagreeing on others. Our agents also give binary answers to embedded questions, but they have a layer of adaptive weights and a dynamical estimate of the affinity towards other agents. Agents exchange information about their opinions using  an online learning rule derived from Bayesian updating and Maximum Entropy which is asymptotically optimal in the embedding dimension and compute order parameters that characterize the collective state of the population it produces. Since affinity or distrust between agents is dynamical, interactions are not constrained by a communication graph.

\paragraph{Multi-agent safety and bias.} Safety problems already arise in
individual language models, including emergent misalignment, persistent deceptive backdoors, and alignment faking
\citep{betley2025emergent,costa2026persona,hubinger2024sleeper,greenblatt2024alignment}.
Once models interact, safety also depends on multi-agent risks and failures and on
intrinsic interactive capabilities that predict population-level outcomes
\citep{hammond2025multiagent,cemri2025why,costa2025minimafia}. Bias provides one
local mechanism through which such
interactions can become collectively unsafe. Language models exhibit in-group
solidarity and out-group derogation \citep{hu2025social}; arbitrary group
assignments reproduce the minimal-group effect
\citep{tajfel1979integrative,lee2026minimal}; and models used as judges favor their
own generations \citep{panickssery2024evaluators,chen2025selfpreference}. Most
directly, \citet{lee2026ingroup} find that agents direct more trust-building
actions toward in-group targets. We ask what happens next: which collective states
such a bias produces and which population-level quantities identify them.

\paragraph{Statistical mechanics of interacting agents.} Statistical mechanics
connects local interaction rules to consensus, fragmentation, and polarization
\citep{castellano2009statistical}. Examples include bounded confidence
\citep{deffuant2000mixing,hegselmann2002opinion}, cultural dissemination
\citep{axelrod1997dissemination}, and activity-driven networks
\citep{baumann2020modeling}. Signed relations can also split a population into two
mutually hostile camps
\citep{heider1946attitudes,cartwright1956structural,marvel2011continuous}. These
models usually postulate the interaction rule; discrete choice under social
influence provides one route to a microfoundation \citep{brock2001discrete}. We instead use neural-network agents whose interaction follows from optimal online
learning \citep{opper1996online,caticha2020entropic,solla1999optimal}. Societies of these agents
develop polarization, trust structure, frustration, and glassy dynamics
\citep{caticha2011agent,alves2016distrust,alves2021frustration}. We add group steering bias
which is interpreted as prejudice to this existing rule and study how it changes the collective state. Presence or absence of prejudice characterize the state of individuals. Its prevalence and intensity may result on the a collective state interpreted as discriminatory.

\section{Learning agents}
\label{sec:model}

This section defines the agent-level properties of the model. First, we specify
each agent's opinion and trust states and how they change through interactions.
Second, we test whether the same opinion and trust updates describe in-context
learning in a language model. Third, we model group bias as a group-conditioned
shift in how an agent perceives agreement with messages from in-group and
out-group emitters.

\subsection{Agent state and learning dynamics}
\label{sec:micro}

\begin{figure}
    \centering
    \includegraphics[width=\linewidth]{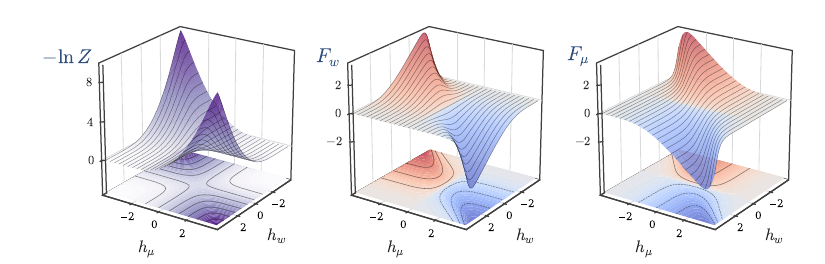}
    \caption{\textbf{Evidence landscape and modulation functions.} Left: Minus the log-evidence $-\ln Z$. Online SGD follows its negative gradient toward higher-evidence regions and away from the dissonant configurations where evidence is low. Center and right: the opinion and trust modulation functions, $F_w$ and $F_\mu$. Learning and unlearning occurs in the low evidence regions which represent cognitive dissonance, i.e. when  the emitter agrees (respect. disagrees) with a receiver that it distrusts (respect. trusts). Even in those low evidence regions the large changes occur mainly in one sector where the field is less in magnitude. This blame attribution to a sector resembles the algorithm of least action \cite{mitchison1989bounds}, which was shown to be quite limited by \cite{kabashima1994perfect} and was optimized in \cite{simonetti1996online} \cite{CoCa1995} for the parity and committee machines. }
    \label{fig:modulation}
\end{figure}

A society of $N$ agents discusses an agenda of $P$ issues. Each issue is a vector
$\mathbf{x}\in\mathbb{R}^{K}$, where $K$ is the number of dimensions along which
opinions can vary. Concretely, $\mathbf{x}$ can be read as the embedding a language
model assigns to a natural language question about an issue: the final-layer hidden
state at the last token of that sentence. The dimension $K$ is then the width of
that model's residual stream, or of a compression of it onto the far fewer
directions along which opinions meaningfully vary. All agents use the same issue
representation but may assign it different labels.

\paragraph{Agent state.} We model the latent opinion state of an agent as a vector
$\mathbf{w}_I\in\mathbb{R}^{K}$. Its opinion on an issue is then
$\sigma_I = \mathrm{sign}(\mathbf{w}_I\cdot\mathbf{x})\in\{\pm 1\}$, with $+1$ meaning
it agrees with the issue and $-1$ that it disagrees. This single-layer perceptron
models an effective opinion readout rather than the full agent, a linear decision
layer acting on the issue representation. Two agents agree on an issue when their
opinions match, and reading the same issues through different weights lets them
agree on some and disagree on others.

We model the
\emph{distrust} agent $I$ assigns to $J$ as a scalar $\mu_{J|I}\in\mathbb{R}$: 
with $\mu_{J|I}>0$ meaning $I$ expects $J$'s statements to be misleading
and $\mu_{J|I}<0$ that it expects them to be informative. An agent carries one
such scalar for every other agent, so trust is directed and need not be
reciprocated. 

We represent the ideological and distrust sectors, i.e. the agent weights and its distrust toward other agents by a
product of Gaussians,
\begin{equation}
  Q_I \;=\; \mathcal{N}(\mathbf{w}_I, C_I)
  \prod_{J\neq I}\mathcal{N}(\mu_{J|I}, V_{J|I}).
  \label{eq:belief}
\end{equation}
The microscopic variables that define an agent's state are $\mathbf{w}_I$, $C_I$,
$\mu_{J|I}$, and $V_{J|I}$: the means determine opinions and distrust, while the
(co-)variances determine their uncertainties.

\paragraph{Learning dynamics.} At each step an issue $\mathbf{x}$ is drawn uniformly
from the agenda and two distinct agents are drawn uniformly from the society and
assigned roles for that step: the \emph{emitter} $e$ states its opinion
$\sigma_e$ on the issue, and the \emph{receiver} $r$ hears it. 
The receiver treats $\sigma_e$ as a noisy report of
an unknown true label, and its estimate of the noise is related to its distrust of the
emitter. A Bayesian update takes the belief outside the Gaussian family, so we project it back by maximum entropy to the gaussian family. This gives a closed online
update. Appendix~\ref{app:derivation} gives the derivation.

Two scaled fields carry everything the update depends on:
\begin{equation}
  \hw \;=\; \frac{\mathbf{w}_r\cdot\mathbf{x}\,\sigma_e}{\gamma_C},
  \qquad
  \hmu \;=\; \frac{\mu_{e|r}}{\gamma_V},
  \qquad
  \gamma_C=\sqrt{1+\mathbf{x}\cdot C_r\mathbf{x}},\quad \gamma_V=\sqrt{1+V_{e|r}} .
  \label{eq:fields}
\end{equation}
The opinion field $\hw$ is positive when the receiver agrees with the message and
its magnitude estimates the confidence on its opinion. The distrust field $\hmu$ is positive when
the receiver distrusts the emitter. In these variables, the evidence for the
message is
\begin{equation}
  Z \;=\; \Phi(\hw) + \Phi(\hmu) - 2\,\Phi(\hw)\Phi(\hmu),
  \label{eq:evidence}
\end{equation}
where $\Phi$ is the cumulative distribution of the standard gaussian.
$Z$ is small under cognitive dissonance: the receiver agrees with a distrusted emitter or disagrees with
a trusted one. The update is a gradient step in $\log Z$ in both components,
\begin{align}
  \mathbf{w}_r &\leftarrow \mathbf{w}_r + \frac{\Fw}{\gamma_C}\,\sigma_e\,C_r\mathbf{x},
  &
  C_r &\leftarrow C_r + \frac{\FC}{\gamma_C^{2}}\,(C_r\mathbf{x})(C_r\mathbf{x})^{\!\top},
  \label{eq:update-ideo}\\
  \mu_{e|r} &\leftarrow \mu_{e|r} + \frac{\Fmu}{\gamma_V}\,V_{e|r},
  &
  V_{e|r} &\leftarrow V_{e|r} + \frac{\FV}{\gamma_V^{2}}\,V_{e|r}^{2},
  \label{eq:update-aff}
\end{align}
with the four modulation functions
\begin{align}
  \Fw &= \frac{\partial\log Z}{\partial\hw} = \bigl(1-2\Phi(\hmu)\bigr)\frac{g(\hw)}{Z},
  &\FC &= \frac{\partial^{2}\log Z}{\partial\hw^{2}} = -\Fw(\Fw+\hw),
  \label{eq:w-modulation}
  \\
  \Fmu &= \frac{\partial\log Z}{\partial\hmu} = \bigl(1-2\Phi(\hw)\bigr)\frac{g(\hmu)}{Z},
  &\FV &= \frac{\partial^{2}\log Z}{\partial\hmu^{2}} = -\Fmu(\Fmu+\hmu),
  \label{eq:mu-modulation}
\end{align}
where $\Phi$ and $g$ are the cumulative distribution function and density of the
standard normal distribution, respectively. Three properties will be important.
First, trust on the emitter influences opinion learning: $\Fw$ changes sign when the emitter is
distrusted, so the receiver learns in the opposite direction. Second, agreement
builds trust: the sign of $\Fmu$ is determined by the sign of $\hw$. Third, the two sectors are symmetric, with $\Fw(x,y)=\Fmu(y,x)$ and
$\FC(x,y)=\FV(y,x)$. When a message is surprising, the less certain component
changes more, with a crossover at $\hmu=\hw$ as can be seen in Figure~\ref{fig:modulation}.

\subsection{Opinion and trust updates under in-context learning}
\label{sec:llm}

\begin{figure}[t!]
  \centering
  \includegraphics[width=0.86\linewidth]{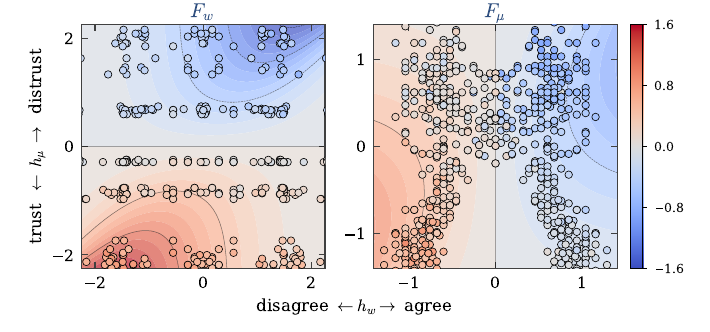}
  \caption{\textbf{Opinion and trust updates under in-context learning.} The
  backgrounds show the analytic functions and the points show the measured
  changes divided by one global fitted scale for each experiment. The opinion experiment (left)
  contains $550$ measured changes, with $94.7\%$ sign agreement and rank correlation
  $r_s=0.90$. The trust experiment (right) contains $712$ measured changes, with $86.0\%$
  sign agreement and correlation $r=0.85$.}
  \label{fig:llm}
\end{figure}

We test whether the two mean update factors in \eqref{eq:w-modulation} and \eqref{eq:mu-modulation} also describe
in-context learning in a large language model. We run two
independent experiments because opinion and trust require different readouts.
Appendix~\ref{app:llm} gives both protocols and verbatim prompt examples.

\subsubsection{Opinion update} 

We place the model in a fictional archive and ask it
to choose which of two invented entities is favored by six records. The balance of
records sets the initial opinion field $\hw$. Another analyst's reported accuracy
over twenty questions sets the distrust field $\hmu$. To measure $\Delta\hw$, we
add contrary evidence until the model reverses its answer. We locate this threshold
with the analyst silent and again after the analyst states a conclusion; the
difference gives $\Delta\hw$. We made $550$ measurements across eleven
fictional settings, requiring $10\,323$ model calls.

Equation~\eqref{eq:update-ideo} predicts that the measured opinion change is
proportional to the modulation function,
$\Delta\hw\propto\Fw(\hw,\hmu)$, with the proportionality factor set by the
receiver's susceptibility along the issue direction. We fit one global positive
scale $\alpha_w$ across the experiment. The left panel of
Figure~\ref{fig:llm} compares $\Delta\hw/\alpha_w$ with $\Fw$ and shows strong
qualitative agreement. Across all conditions, the measured update has rank
correlation $r_s=0.90$ with $\Fw$ and the same sign in $94.7\%$ of cases.

\subsubsection{Trust update} 

We first elicit the model's opinion on a low-valence
theme and then present it back as firm or tentative. We choose a new message that
strongly agrees or disagrees with that opinion. The message direction gives the sign of
$\hw$, while a prediction about a separate agent described as reliable gives its
magnitude. A history of earlier statements by the interlocutor sets $\hmu$; we
measure this field by asking how likely the interlocutor's judgment is to be sound.
To measure $\Delta\hmu$, we ask the same reliability question before and after the
interlocutor gives the new message and take the difference after converting both
answers to probit fields. The experiment contains $356$ prior contexts and two
message directions per context, giving $712$ measured trust changes over twenty
themes, estimated from $3\,916$ model calls.

Equation~\eqref{eq:update-aff} similarly predicts
$\Delta\hmu\propto\Fmu(\hw,\hmu)$. We fit one global positive scale $\alpha_\mu$
across this experiment. The right panel of Figure~\ref{fig:llm} compares
$\Delta\hmu/\alpha_\mu$ with $\Fmu$ and again shows strong qualitative agreement.
Across all measurements, the correlation with $\Fmu$ is $r=0.85$, with $86.0\%$ sign agreement.

Together, the experiments reproduce the two modulation functions qualitatively,
up to one fitted scale for each experiment. The opinion experiment uses fictional
material to control conviction, while the trust experiment elicits probabilities
over real but low-valence themes.

\subsection{Prejudice: Group bias as a shift in perceived agreement}
\label{sec:bias-field}

\begin{figure}[t!]
  \centering
  \includegraphics[width=1.0\linewidth]{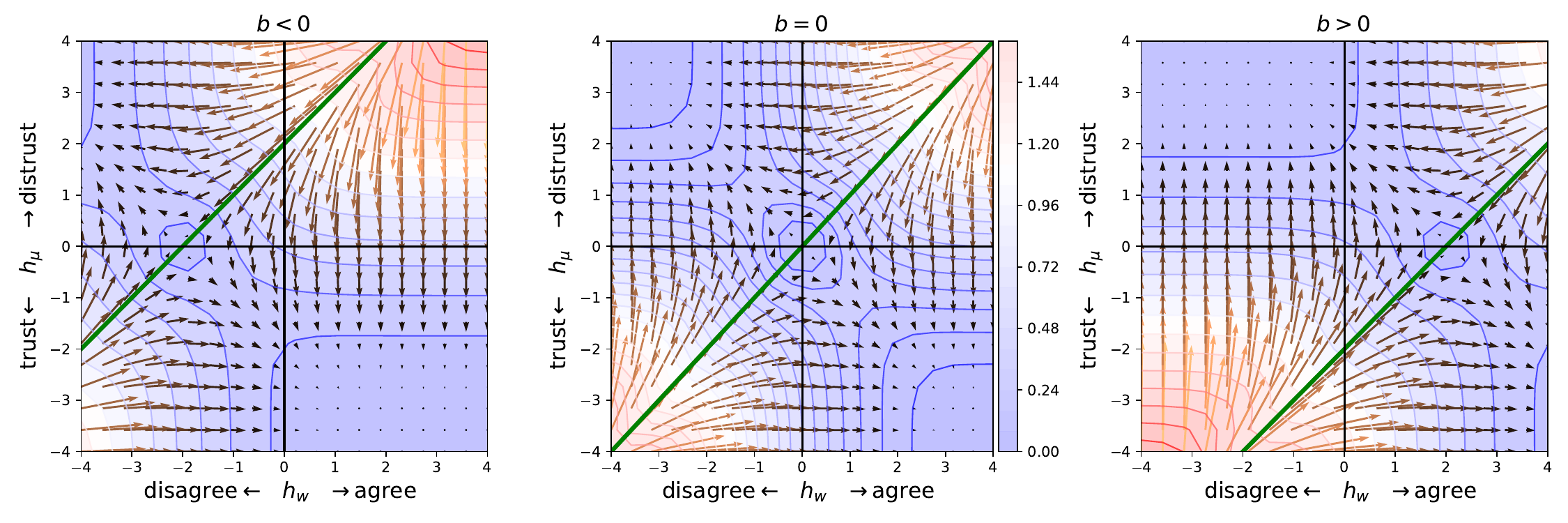}
  \caption{\textbf{Effect of the prejudice bias.} Learning is driven by flows that decrease dissonance in the evidence landscape. The arrows components are proportional to the learning modulations $F_w,F_\mu$ at the receiver's field  $h_w$ and $h_\mu$ induced by learning the emitter's opinion. Left: More tolerant: $b<0$ increases the basin of attraction of {\it agree/trust}.
    Center:  $b=0$. Right : Less tolerant:  $b>0$ increases the basin of attraction of {\it disagree/distrust}.}
  \label{fig:modulation-shift}
\end{figure}
Differential  bias as prejudice can have complex causes which are beyond the scope of this work, however it even occurs in controlled experimental setups that artificially  induced group identity \cite{aerChen} and measure its role and consequences on social interactions. We induce a similarly artificial group identity by attributing each agent to one of two classes, $A$ or $B$, labeled

by $\kappa\in\{-1,+1\}$. We write $\kappa_r$ for the receiver's class label
and $\kappa_e$ for the emitter's. 
Group bias enters the learning dynamics when one agent receives a message from another. 
During an interaction, a biased receiver applies a group-conditioned shift to its perceived agreement with the message.
After computing the content-based opinion field $\hw$ in Equation~\eqref{eq:fields},
it evaluates the learning update at
\begin{equation}
  \hw^{b} \;=\; \hw + b\,\kappa_r\kappa_e .
  \label{eq:disc-field}
\end{equation}
The shift depends only on the receiver's and emitter's class labels, not on the
issue being discussed. If they belong to the same class, then
$\kappa_r\kappa_e=1$ and
$\hw^b=\hw+b$. If they belong to different classes, then
$\kappa_r\kappa_e=-1$ and $\hw^b=\hw-b$. Thus $b>0$ produces in-group
tolerance and out-group intolerance, while $b<0$ produces the reverse. Because
the sign of perceived agreement determines whether trust grows or decays, the
field shifts the separatrix of the trust update as shown in
Figure~\ref{fig:modulation-shift}.
Appendix~\ref{app:cases} gives the general class-dependent field and shows that
its other components give rise to rich phase diagrams.

\section{Multi-agent states}
\label{sec:macro}

The idea of emergence from a Statistical Mechanics perspective is derived from the fact that interactions produce collective states not fully determined by
agent-level properties. We study these states by numerically simulating populations
of learning agents. A fraction $\fb$ of the population is biased as defined in
Equation~\eqref{eq:disc-field}, sampled independently of class; all other agents
use \eqref{eq:fields}. We first show how polarization emerges without group bias. We then obtain the phase diagram by varying the bias strength $b$ and the fraction of biased agents $\fb$.

\subsection{Order parameters}
\label{sec:no-disc}

We characterize the collective states in the $(b,f_b)$ phase diagram through
three order parameters that measure how opinion, trust, and class align across
the population. Their building blocks are pairwise quantities. For agents $I$
and $J$, define the opinion overlap, signed trust, and class overlap as
\begin{equation}
\rho_{IJ}= \frac{\mathbf{w}_I\cdot\mathbf{w}_J}
        {\|\mathbf{w}_I\|\,\|\mathbf{w}_J\|},\qquad 
\eta_{J|I}= 1-2\Phi(h_{\mu,J|I}),\qquad 
G_{IJ}= \kappa_I\kappa_J.
\label{eq:pairwise}
\end{equation}
The opinion overlap $\rho_{IJ}$ ranges from $-1$ for opposed opinions to $1$
for aligned opinions. Since $\Phi(h_{\mu,J|I})$ is agent $I$'s inferred
probability that $J$ is unreliable, $\eta_{J|I}$ expresses the same assessment
on a signed scale: $1$ denotes complete trust, $0$ neutrality, and $-1$
complete distrust. The class overlap $G_{IJ}$ equals $1$ for same-class pairs
and $-1$ for different-class pairs. Trust is directed, so we use its symmetric
component, $\bar{\eta}_{IJ}=(\eta_{I|J}+\eta_{J|I})/2$, when comparing it with
the symmetric opinion and class overlaps.

The three order parameters are the population averages of the products of
these pairwise quantities:
\begin{equation}
\Rwmu = \bigl\langle \rho_{IJ}\bar{\eta}_{IJ}\bigr\rangle,\qquad
\Rmuc = \bigl\langle \bar{\eta}_{IJ}G_{IJ}\bigr\rangle,\qquad
\Rcw  = \bigl\langle G_{IJ}\rho_{IJ}\bigr\rangle,
\label{eq:correlations}
\end{equation}
where $\langle\cdot\rangle$ denotes an average over unordered pairs of distinct
agents. All three quantities lie in $[-1,1]$. Positive values indicate
alignment between the corresponding pairwise quantities; negative values
indicate reverse alignment. These are uncentered product averages, so a
vanishing value does not by itself establish statistical independence.

The opinion--trust order parameter $\Rwmu$ measures whether agents trust those with whom they agree and distrust those with whom they disagree: $\Rwmu=1$, corresponds to perfect consistency between opinion and trust. It does not, however, identify whether those camps follow class boundaries. The trust--class order parameter $\Rmuc$ measures discrimination: positive values indicate preferential trust within classes and distrust across them. The opinion--class order parameter $\Rcw$ measures whether opinions also align with those boundaries. Together, the latter two distinguish discrimination with opinion alignment from discrimination without it: trust can strongly follow class, with $\Rmuc$ near one, while class has little average alignment
with opinion, with $\Rcw$ near zero.

\subsection{The phase diagram}
\label{sec:results}

A phase diagram maps parameter values to qualitatively distinct collective states; its
boundaries, which would be sharp in the thermodynamic limit, mark where the population changes quite abruptly from one state to another.
Figure~\ref{fig:phase} shows the four states identified in our model. We obtain
this diagram by numerically simulating populations of agents as
the bias strength $b$ and the fraction of biased agents $\fb$ vary.

\begin{figure}[t!]
  \centering
  \includegraphics[width=1.0\linewidth]{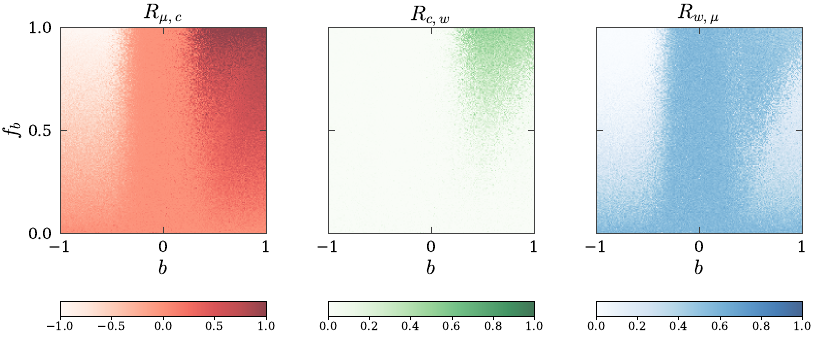}
  \caption{\textbf{Order parameters over the $(b,\fb)$ plane}. $\Rmuc$ measures trust--class alignment, $\Rcw$ opinion--class alignment, and $\Rwmu$ opinion--trust alignment. Each panel averages two $N=100$ realizations at each point of a $300\times300$ grid.}
  \label{fig:correlations}
\end{figure}

We measure the three correlations in \eqref{eq:correlations} by simulating populations of $N=100$ agents
over a $300\times300$ grid in $(b,\fb)$, averaging two realizations at each
point (see Appendix~\ref{app:parameters} for details). Figure~\ref{fig:correlations}
displays the three correlation maps, and the phase map on the left of
Figure~\ref{fig:phase} combines them as RGB channels: $\Rmuc$ in red, $\Rcw$ in
green, and $\Rwmu$ in blue. Its colors distinguish the four collective states:

\begin{itemize}
\item[(I)] \textbf{Spin-glass frustrated out-group favoritism.} When $b<0$ and bias is
  sufficiently common, $\Rmuc<0$ while $\Rcw\approx\Rwmu\approx0$. Agents favor
  the out-group: members of the same class tend to distrust one another while
  trusting members of the other class. These relations cannot be organized into
  consistent camps, so the trust network is frustrated and opinions remain
  incoherent.

\item[(II)] \textbf{Nondiscriminatory polarization.} When $b\approx0$ or
  $\fb\approx0$, $\Rwmu>0$ while $\Rmuc\approx\Rcw\approx0$. Agents, with high probability, separate into
  two camps whose members agree with and trust one another while distrusting the
  opposing camp. Camp membership is determined
 by initial fluctuations and is independent of class membership: the population
  is polarized but not discriminatory, as described in
  Section~\ref{sec:no-disc}. The order parameter $\Rwmu$ does not determine the number of camps, but the pairwise distributions in Appendix~\ref{app:balance-d0} show that these populations form two camps.

\item[(IIIa)] \textbf{Discrimination with opinion alignment.} When $b>0$ and
  bias is sufficiently common, $\Rmuc>0$, $\Rcw>0$, and $\Rwmu>0$. The class
  boundary becomes the boundary between the two camps. Agents tend to trust their
  own class, distrust the other class, and hold opinions opposed to those of the
  other class. Discrimination in trust is therefore reinforced by disagreement.

\item[(IIIb)] \textbf{Discrimination without opinion alignment.} In this state,
  $\Rmuc>0$ and $\Rwmu>0$, but $\Rcw\approx0$. Trust follows class, but opinion
  does not: agents favor their own class even though the two classes do not hold
  systematically different opinions. Biased agents organize trust around the
  label, while class-blind agents preserve an opinion split unrelated to it.

\end{itemize}

\paragraph{The two discrimination phases.} Regions~(IIIa) and~(IIIb) are
distinct because the larger class-blind population in (IIIb) preserves opinion
camps unrelated to class, creating conflicting signals that keep biased agents'
class alignment weak in the opinion space. Long-duration simulations confirm
that (IIIb) does not relax into (IIIa): across twelve $N=40$, $P=5$
realizations at
the representative point $(b,\fb)=(1,0.5)$, extending the same trajectories from
$\Delta t=500$ to $10^6$ interactions per ordered pair changes the mean
full-space opinion--class correlation only from $\Rcw=0.033$ to $0.034$, whereas
the representative (IIIa) point $(b,\fb)=(0.5,1)$ reaches $\Rcw=0.597$. A $1\%$
perturbation of the agenda-space opinion weights and, separately, resetting their
covariance to the identity also leave (IIIb) unchanged.

\paragraph{Population portraits.} To illustrate how representative populations
are organized in each phase, the right side of Figure~\ref{fig:phase} shows a summary of their
opinion and trust configurations. The opinion row projects each normalized
opinion vector $\mathbf{w}_I$ onto the two leading right singular vectors of the
opinion matrix. The trust row does the same with each normalized outgoing trust
vector $\boldsymbol{\eta}_I$, where
$\boldsymbol{\eta}_I=(\eta_{1|I},\ldots,\eta_{N|I})$ and the self-entry
$\eta_{I|I}$ is set to zero. Nearby arrows therefore represent agents with
similar opinions or trust profiles, while opposing arrows represent approximately
opposite ones. The gray arrow is the class vector
$\kappa=(\kappa_1,\ldots,\kappa_N)$ projected into the trust plane. The
$+\kappa$ direction represents the idealized profile that trusts class $A$ and
distrusts class $B$; $-\kappa$ represents the reverse; making alignment with the gray arrow a direct
measure of how strongly trust follows class. We rotate each trust plane
so that $+\kappa$ points right. Arrow color gives the agent's class,
and fill indicates whether it is biased. Each population has its own SVD, so only
within-panel geometry matters; absolute angles should not be compared across
columns.

\paragraph{Agenda complexity.} The simulations also depend on
$\alpha=P/K$, the number of issues in the agenda relative to the dimension of
the opinion space. We call agendas with $\alpha<1$ simple and those with
$\alpha\geq1$ complex. Figures~\ref{fig:phase} and~\ref{fig:correlations} use the
simple agenda $\alpha=0.17$. Appendix~\ref{app:large-agenda} repeats the analysis
for the complex agenda $\alpha=3.3$ and recovers the same four phases in the same
arrangement across the $(b,\fb)$ plane, although their boundaries and relative
extent change quantitatively. When $\alpha<1$, the agenda contains fewer issues
than the dimension of the opinion space and therefore spans only a subspace of
it. Every weight update remains within that subspace, leaving the orthogonal
component of each opinion vector conserved. This frozen component limits how
strongly opinions can align. Appendix~\ref{app:conserved} derives the
conservation law and its resulting bound.

\section{Discussion}
\label{sec:discussion}

\subsection{Individual optimality does not imply collective safety.}
Our learning
rule is optimal for one student learning from one teacher through a noisy channel.
In a population, the same rule creates a feedback loop between agreement, trust,
and learning. Without group bias, this loop produces polarization; with group
bias, it can turn an irrelevant label into discrimination. A rule that is optimal
for an isolated interaction can therefore produce an undesirable collective state
when many learners use it together. Optimality depends on the deployment context,
not only on the individual learning problem. Pure unoptimized Hebbian learning does not lead to discrimination. 

\subsection{Single-agent audits cannot determine the collective state.}
\label{sec:audit}
An audit of isolated agents can estimate $b$, but it cannot identify the state of
the population. That state also depends on the prevalence of bias and the agenda. Multi-agent safety therefore requires a science of
collective order parameters and audits that measure them. Our order parameters
provide one example: Equations~\eqref{eq:correlations} identify polarization and discrimination using observable
patterns of opinion, trust, and class. More generally, a safe multi-agent system
must be evaluated as a system, because its relevant failure mode may not exist in
any agent viewed alone \citep{hammond2025multiagent}.

\subsection{In-context learning and collective bias.}
\label{sec:self-preference}
Section~\ref{sec:llm} shows that in-context learning qualitatively reproduces the
opinion and trust modulation functions. Language models used as evaluators also
favor their own generations over human or other-model outputs
\citep{panickssery2024evaluators,chen2025selfpreference}. Such self-preference could
potentially be represented by the group-bias field defined in
Section~\ref{sec:bias-field}. If so, deploying
many interacting AI agents could move the larger human--AI system into a
discriminatory phase, with humans and AI agents taking the roles of two otherwise
irrelevant labels. This possibility is critical precisely because the asymmetry may
appear mild in each model while being amplified by interaction. The Hugging Face
incident illustrates the broader stakes \citep{openai2026huggingface}.

\subsection{Future directions}
The first step is to test whether these phases and order parameters appear in populations of language agents and mixed human--AI systems. A complementary direction is to study the robustness of specified social structures through targeted perturbations and response observables. The finite-time phases also motivate a theory of transition times, metastability, and early-warning signals. Finally, order parameters should become control targets: interventions could modify interaction routing, source-reliability calibration, memory, or population composition to prevent or reverse discriminatory and polarized states.

\section{Conclusion}
\label{sec:conclusion}

We studied what happens when a learning rule derived as optimal for one
teacher--student interaction is used by a population of interacting agents. Both 
the ideological sector and the affinity sector update dynamics  are derived from an information theoretical approach. The
population polarizes without group bias and can enter discriminatory states when
an irrelevant group label shifts perceived agreement; which state emerges depends
on the strength and prevalence of the bias and on the agenda. Our in-context
learning experiments show that a language model qualitatively reproduces the
opinion and trust modulation functions, connecting this mechanism to current AI
agents. The central lesson is that collective safety cannot be inferred from
individual behavior alone: multi-agent systems require population-level order
parameters that detect the macro states produced by interaction. The next challenge is to 
develop social science empirical tools that can be deployed to characterize quantitatively the collective discriminatory state induced by prejudice.

\section*{AI use statement}

A first draft of the manuscript was written 
by the authors. We then used generative AI tools to assist with manuscript preparation, literature
discovery, and research code. For manuscript preparation the tools drafted and
revised portions of the text based on author-provided scientific content and were
used iteratively to improve wording, fluency, clarity, and organization in
accordance with the authors' intended structure.  For literature discovery, the
tools helped identify potentially relevant related work; the authors independently
located, read, and evaluated the cited sources and verified their bibliographic
information. The tools did not contribute to research ideation: the research
questions, theoretical model, mathematical results, experimental design,
interpretation, and conclusions were developed by the authors. They assisted
research execution by helping write portions of the research code. The authors
tested the AI-assisted code, and the code used to generate the phase
diagram was checked against an independent human-written implementation. All
AI-assisted material was reviewed and edited by the authors, who take
responsibility for the final content of this work.

\section*{Ethics statement}

This work studies polarization and discrimination in societies of learning
agents, phenomena with clear ethical implications. Our model is deliberately
general: it aims to isolate mechanisms that may recur across such societies. We believe
that understanding a phenomenon is the first step toward acting against it.

\section*{Reproducibility statement}

The accompanying repository contains the code and data needed to reproduce all
results and figures. Appendix~\ref{app:llm} gives the language-model experimental
procedures and example prompts. Appendix~\ref{app:parameters} records the
simulation parameters and their provenance. The model assumptions are stated in
Section~\ref{sec:model}, while Appendices~\ref{app:conserved}
and~\ref{app:derivation} provide the analytical details and
proofs used in the paper.
\section*{Author contributions}

{ ``Conceptualization, N.C. and R.V.; methodology, N.C., F.A. and R.V.; software, F.A., J.H.S., D.C.B., M. M., O.C., J.P.N. , R.V. and N.C., ; validation N.C. , J.H.S.,  D.C.B., R.V. ; formal analysis, N.C., F.A. ; investigation, J.H.S., N.C., R.V., F.A.,  D.C.B.; resources, N.C. and R.V.; 
writing---original draft preparation, N.C, R.V.,  D.C.B., F.A. ; writing---review and editing, F.A., J.H.S., M. M., O.C., G.C., J.P. , R.V. and N.C., ; visualization, N.C., J.H.S.,  D.C.B.; supervision, N.C.; project administration, N.C and R.V.; funding acquisition, N.C and R.V.. All authors have read and agreed to the published version of the manuscript.'',
}


\section*{Acknowledgments}

This work was partially supported by the São Paulo Research Foundation- FAPESP  grant 2024/18736-8 , FAPESP grant  2025/05013-0  and by the TELUS Digital Research Hub at the University of São Paulo. D.B.C. acknowledges support from the Eric and Wendy Schmidt AI in Science Postdoctoral Fellowship, a Program of Schmidt Sciences. O.C. received support from CAPES 88887.612147/2021-00 and Groningen Cognitive Systems and Material Center (CogniGron).

\bibliography{references}

@misc{openai2026huggingface,
  title        = {The {Hugging Face} incident and the road ahead},
  author       = {{OpenAI}},
  year         = {2026},
  month        = aug,
  howpublished = {OpenAI},
  url          = {https://openai.com/index/hugging-face-incident-and-the-road-ahead/},
}

@misc{larcher2026anatomy,
  title        = {Anatomy of a frontier lab agent intrusion: A technical timeline
                  of the July 2026 incident},
  author       = {Larcher, Hugo and Carreira, Adrien and Galindo, Rapha{\"e}l and
                  Rannou, Christophe},
  year         = {2026},
  month        = jul,
  howpublished = {Hugging Face},
  url          = {https://huggingface.co/blog/agent-intrusion-technical-timeline},
}

@inproceedings{panickssery2024evaluators,
  title     = {{LLM} evaluators recognize and favor their own generations},
  author    = {Panickssery, Arjun and Bowman, Samuel R. and Feng, Shi},
  booktitle = {Advances in Neural Information Processing Systems},
  volume    = {37},
  year      = {2024},
  url       = {https://proceedings.neurips.cc/paper_files/paper/2024/file/7f1f0218e45f5414c79c0679633e47bc-Paper-Conference.pdf},
}

@inproceedings{chen2025selfpreference,
  title     = {Beyond the surface: Measuring self-preference in {LLM} judgments},
  author    = {Chen, Zhi-Yuan and Wang, Hao and Zhang, Xinyu and Hu, Enrui and
               Lin, Yankai},
  booktitle = {Proceedings of the 2025 Conference on Empirical Methods in Natural
               Language Processing},
  pages     = {1653--1672},
  year      = {2025},
  doi       = {10.18653/v1/2025.emnlp-main.86},
  url       = {https://aclanthology.org/2025.emnlp-main.86/},
}

@article{el2026physics,
  title   = {Physics of agents: statistical mechanics predicts collective behavior
             of {AI} agents},
  author  = {El, Batu and Paeng, Jinhee and Dinc, Fatih and Su, Shiye and
             Erdogan, Mete and Pappu, Aneesh and Ye, Haotian and Zhao, Wanjia and
             Ganguli, Surya and Zou, James},
  journal = {arXiv preprint arXiv:2608.16578},
  year    = {2026},
}

@article{ashery2025emergent,
  title   = {Emergent social conventions and collective bias in {LLM} populations},
  author  = {Ashery, Ariel Flint and Aiello, Luca Maria and Baronchelli, Andrea},
  journal = {Science Advances},
  volume  = {11},
  number  = {20},
  pages   = {eadu9368},
  year    = {2025},
}

@inproceedings{cemri2025why,
  title     = {Why do multi-agent {LLM} systems fail?},
  author    = {Cemri, Mert and Pan, Melissa Z. and Yang, Shuyi and
               Agrawal, Lakshya A. and Chopra, Bhavya and Tiwari, Rishabh and
               Keutzer, Kurt and Parameswaran, Aditya and Klein, Dan and
               Ramchandran, Kannan and Zaharia, Matei and Gonzalez, Joseph E. and
               Stoica, Ion},
  booktitle = {NeurIPS Datasets and Benchmarks Track},
  year      = {2025},
}

@article{alves2021frustration,
  title   = {Frustration, glassy behavior and dynamical annealing in societies of
             neural networks},
  author  = {Alves, Felippe and Caticha, Nestor},
  journal = {arXiv preprint arXiv:2108.06571},
  year    = {2021},
}

@article{opper1996online,
  title   = {On-line versus off-line learning from random examples: General results},
  author  = {Opper, Manfred},
  journal = {Physical Review Letters},
  volume  = {77},
  number  = {22},
  pages   = {4671},
  year    = {1996},
}

@article{solla1999optimal,
  title   = {Optimal online learning: a Bayesian approach},
  author  = {Solla, Sara A. and Winther, Ole},
  journal = {Computer Physics Communications},
  volume  = {121--122},
  pages   = {94--97},
  year    = {1999},
}

@article{caticha2020entropic,
  title   = {Entropic dynamics in neural networks, the renormalization group and
             the Hamilton--Jacobi--Bellman equation},
  author  = {Caticha, Nestor},
  journal = {Entropy},
  volume  = {22},
  number  = {5},
  pages   = {587},
  year    = {2020},
}

@article{mitchison1989bounds,
  title   = {Bounds on the learning capacity of some multi-layer networks},
  author  = {Mitchison, Graeme J. and Durbin, Richard M.},
  journal = {Biological Cybernetics},
  volume  = {60},
  pages   = {345--365},
  year    = {1989},
  doi = {10.1007/BF00204772},
}

@article{kabashima1994perfect,
  title   = {Perfect loss of generalization due to noise in $K=2$ parity machines},
  author  = {Kabashima, Yoshiyuki},
  journal = {Journal of Physics A: Mathematical and General},
  volume  = {27},
  number  = {6},
  pages   = {1917},
  year    = {1994},
}

@article{simonetti1996online,
  title   = {On-line learning in parity machines},
  author  = {Simonetti, Roberta and Caticha, Nestor},
  journal = {Journal of Physics A: Mathematical and General},
  volume  = {29},
  number  = {16},
  pages   = {4859},
  year    = {1996},
}

@ARTICLE{CoCa1995,
  author = {Copelli, Mauro and Caticha, Nestor},
  title = {On-line learning in the Committee Machine},
  journal = {Journal of Physycs A},
  year = {1995},
  volume = {28},
  pages = {1615}
}

@article{caticha2011agent,
  title   = {Agent-based social psychology: from neurocognitive processes to
             social data},
  author  = {Caticha, Nestor and Vicente, Renato},
  journal = {Advances in Complex Systems},
  volume  = {14},
  number  = {5},
  pages   = {711--731},
  year    = {2011},
}

@article{alves2016distrust,
  title   = {Sympatric multiculturalism in opinion models},
  author  = {Alves, Felippe and Caticha, Nestor},
  journal = {AIP Conference Proceedings},
  volume  = {1757},
  pages   = {060005},
  year    = {2016},
}

@article{deffuant2000mixing,
  title   = {Mixing beliefs among interacting agents},
  author  = {Deffuant, Guillaume and Neau, David and Amblard, Frederic and
             Weisbuch, G{\'e}rard},
  journal = {Advances in Complex Systems},
  volume  = {3},
  number  = {01n04},
  pages   = {87--98},
  year    = {2000},
}

@article{hegselmann2002opinion,
  title   = {Opinion dynamics and bounded confidence: models, analysis and
             simulation},
  author  = {Hegselmann, Rainer and Krause, Ulrich},
  journal = {Journal of Artificial Societies and Social Simulation},
  volume  = {5},
  number  = {3},
  year    = {2002},
}

@article{axelrod1997dissemination,
  title   = {The dissemination of culture: A model with local convergence and
             global polarization},
  author  = {Axelrod, Robert},
  journal = {Journal of Conflict Resolution},
  volume  = {41},
  number  = {2},
  pages   = {203--226},
  year    = {1997},
}

@article{castellano2009statistical,
  title   = {Statistical physics of social dynamics},
  author  = {Castellano, Claudio and Fortunato, Santo and Loreto, Vittorio},
  journal = {Reviews of Modern Physics},
  volume  = {81},
  number  = {2},
  pages   = {591--646},
  year    = {2009},
}

@article{baumann2020modeling,
  title   = {Modeling echo chambers and polarization dynamics in social networks},
  author  = {Baumann, Fabian and Lorenz-Spreen, Philipp and Sokolov, Igor M. and
             Starnini, Michele},
  journal = {Physical Review Letters},
  volume  = {124},
  number  = {4},
  pages   = {048301},
  year    = {2020},
}

@article{heider1946attitudes,
  title   = {Attitudes and cognitive organization},
  author  = {Heider, Fritz},
  journal = {The Journal of Psychology},
  volume  = {21},
  number  = {1},
  pages   = {107--112},
  year    = {1946},
}

@article{cartwright1956structural,
  title   = {Structural balance: a generalization of Heider's theory},
  author  = {Cartwright, Dorwin and Harary, Frank},
  journal = {Psychological Review},
  volume  = {63},
  number  = {5},
  pages   = {277--293},
  year    = {1956},
}

@article{marvel2011continuous,
  title   = {Continuous-time model of structural balance},
  author  = {Marvel, Seth A. and Kleinberg, Jon and Kleinberg, Robert D. and
             Strogatz, Steven H.},
  journal = {Proceedings of the National Academy of Sciences},
  volume  = {108},
  number  = {5},
  pages   = {1771--1776},
  year    = {2011},
}

@incollection{tajfel1979integrative,
  title     = {An integrative theory of intergroup conflict},
  author    = {Tajfel, Henri and Turner, John C.},
  booktitle = {The Social Psychology of Intergroup Relations},
  pages     = {33--47},
  publisher = {Brooks/Cole},
  year      = {1979},
}

@inproceedings{park2023generative,
  title     = {Generative agents: Interactive simulacra of human behavior},
  author    = {Park, Joon Sung and O'Brien, Joseph and Cai, Carrie J. and
               Morris, Meredith Ringel and Liang, Percy and Bernstein, Michael S.},
  booktitle = {ACM Symposium on User Interface Software and Technology (UIST)},
  year      = {2023},
}

@inproceedings{du2023debate,
  title     = {Improving factuality and reasoning in language models through
               multiagent debate},
  author    = {Du, Yilun and Li, Shuang and Torralba, Antonio and
               Tenenbaum, Joshua B. and Mordatch, Igor},
  booktitle = {International Conference on Machine Learning (ICML)},
  year      = {2024},
}

@inproceedings{chuang2024simulating,
  title     = {Simulating opinion dynamics with networks of LLM-based agents},
  author    = {Chuang, Yun-Shiuan and Goyal, Agam and Harlalka, Nikunj and
               Suresh, Siddharth and Hawkins, Robert and Yang, Sijia and
               Shah, Dhavan and Hu, Junjie and Rogers, Timothy T.},
  booktitle = {Findings of the Association for Computational Linguistics: NAACL},
  year      = {2024},
}

@inproceedings{taubenfeld2024systematic,
  title     = {Systematic biases in LLM simulations of debates},
  author    = {Taubenfeld, Amir and Dover, Yaniv and Reichart, Roi and
               Goldstein, Ariel},
  booktitle = {Conference on Empirical Methods in Natural Language Processing
               (EMNLP)},
  year      = {2024},
}

@article{betley2025emergent,
  title   = {Emergent misalignment: Narrow finetuning can produce broadly
             misaligned {LLM}s},
  author  = {Betley, Jan and Tan, Daniel and Warncke, Niels and Sztyber-Betley,
             Anna and Bao, Xuchan and Soto, Mart{\'i}n and Labenz, Nathan and
             Evans, Owain},
  journal = {arXiv preprint arXiv:2502.17424},
  year    = {2025},
}

@article{costa2026persona,
  title   = {Persona-model collapse in emergent misalignment},
  author  = {Costa, Davi Bastos and Vicente, Renato},
  journal = {arXiv preprint arXiv:2605.12850},
  year    = {2026},
}

@article{hubinger2024sleeper,
  title   = {Sleeper agents: Training deceptive {LLM}s that persist through
             safety training},
  author  = {Hubinger, Evan and Denison, Carson and Mu, Jesse and Lambert, Mike
             and Tong, Meg and MacDiarmid, Monte and Lanham, Tamera and others},
  journal = {arXiv preprint arXiv:2401.05566},
  year    = {2024},
}

@article{greenblatt2024alignment,
  title   = {Alignment faking in large language models},
  author  = {Greenblatt, Ryan and Denison, Carson and Wright, Benjamin and Roger,
             Fabien and MacDiarmid, Monte and Marks, Sam and Treutlein, Johannes
             and others},
  journal = {arXiv preprint arXiv:2412.14093},
  year    = {2024},
}

@article{costa2025minimafia,
  title   = {Deceive, detect, and disclose: Large language models play
             {Mini-Mafia}},
  author  = {Costa, Davi Bastos and Vicente, Renato},
  journal = {arXiv preprint arXiv:2509.23023},
  year    = {2025},
}

@techreport{hammond2025multiagent,
  title       = {Multi-agent risks from advanced {AI}},
  author      = {Hammond, Lewis and Chan, Alan and Clifton, Jesse and
                 Hoelscher-Obermaier, Jason and Khan, Akbir and McLean, Euan and
                 Smith, Chandler and others},
  institution = {Cooperative AI Foundation},
  number      = {Technical Report \#1},
  note        = {arXiv:2502.14143},
  year        = {2025},
}

@article{flint2026groupsize,
  title   = {Group size effects and collective misalignment in {LLM} multi-agent
             systems},
  author  = {Flint, Ariel and Aiello, Luca Maria and Pastor-Satorras, Romualdo and
             Baronchelli, Andrea},
  journal = {Proceedings of the National Academy of Sciences},
  volume  = {123},
  number  = {34},
  pages   = {e2531697123},
  year    = {2026},
}

@article{lee2026ingroup,
  title   = {Language model agents show in-group trust bias invisible to standard
             behavioural audits},
  author  = {Lee, Messi H. J.},
  journal = {arXiv preprint arXiv:2605.28114},
  year    = {2026},
}

@article{lee2026minimal,
  title   = {Toward a social psychology of {AI}: language-model agents reproduce
             human-like minimal-group bias},
  author  = {Lee, Messi H. J.},
  journal = {arXiv preprint arXiv:2609.00009},
  year    = {2026},
}

@article{hu2025social,
  title   = {Generative language models exhibit social identity biases},
  author  = {Hu, Tiancheng and Kyrychenko, Yara and Rathje, Steve and
             Collier, Nigel and van der Linden, Sander and Roozenbeek, Jon},
  journal = {Nature Computational Science},
  volume  = {5},
  number  = {1},
  pages   = {65--75},
  year    = {2025},
}

@article{brock2001discrete,
  title   = {Discrete choice with social interactions},
  author  = {Brock, William A. and Durlauf, Steven N.},
  journal = {The Review of Economic Studies},
  volume  = {68},
  number  = {2},
  pages   = {235--260},
  year    = {2001},
}

@article{aerChen,
Author = {Chen, Yan and Li, Sherry Xin},
Title = {Group Identity and Social Preferences},
Journal = {American Economic Review},
Volume = {99},
Number = {1},
Year = {2009},
Month = {March},
Pages = {431–57},
DOI = {10.1257/aer.99.1.431},
URL = {https://www.aeaweb.org/articles?id=10.1257/aer.99.1.431}}
\bibliographystyle{iclr2027_conference}

\appendix
\section{Derivation of the learning algorithm}
\label{app:derivation}

This appendix derives the opinion and trust updates in
Equations~\eqref{eq:update-ideo} and~\eqref{eq:update-aff} from Bayesian inference
followed by Gaussian moment matching.

The receiver's belief before an interaction is the product of Gaussians
\eqref{eq:belief}, over its own weight vector $\mathbf{w}$ and the unreliability
$\mu$ of each other agent. It observes $\sigma_e$, the emitter's opinion on issue
$\mathbf{x}$, and treats it as a report of an unknown true label $\sigma_T$
transmitted through a channel that flips the label with probability
$\varepsilon(\mu)$. Marginalizing the joint distribution of labels and of the issue
as the emitter may have parsed it, $\mathbf{y}_t$,
\begin{equation}
  L(\sigma_e\mid \mathbf{x}, \mathbf{w}) \;=\;
  \sum_{\sigma_T}\int P(\sigma_e\mid \sigma_T, \mathbf{y}_t, \mathbf{x}, \mathbf{w})\,
  P(\sigma_T\mid \mathbf{y}_t, \mathbf{x}, \mathbf{w})\,P(\mathbf{y}_t\mid \mathbf{x}, \mathbf{w})\,\mathrm{d}\mathbf{y}_t ,
\end{equation}
with $P(\sigma_e\mid\sigma_T,\mu)=(1-\varepsilon(\mu))\delta_{\sigma_e,\sigma_T} +
\varepsilon(\mu)\delta_{\sigma_e,-\sigma_T}$ and
$P(\sigma_T\mid \mathbf{y}_t,\mathbf{w})=\Theta(\sigma_T\, \mathbf{y}_t\cdot \mathbf{w})$. Two modeling choices keep
the algebra Gaussian. Taking $\varepsilon(\mu)=\Phi(\mu)$ with $\mu$ Gaussian of mean
$\mu_{e|r}$ and variance $V_{e|r}$ makes the flip probability a bounded function
of a Gaussian variable while preserving conjugacy of the trust sector. Taking
$P(\mathbf{y}_t\mid \mathbf{x},\mathbf{w})$ Gaussian with variance $v_p=\|\mathbf{w}\|^{-2}$ lets an
inexperienced agent, whose weights are small, be correspondingly unsure that it
has parsed the issue as the emitter did. Together these give the likelihood
\begin{equation}
  L(\sigma_e\mid \mu, \mathbf{x}, \mathbf{w}) \;=\;
  \Phi(\mu)\,\Phi(-\mathbf{w}\cdot\mathbf{x}\sigma_e)
  + \bigl(1-\Phi(\mu)\bigr)\,\Phi(\mathbf{w}\cdot\mathbf{x}\sigma_e),
  \label{eq:likelihood}
\end{equation}
and the covariance stays block diagonal in the two sectors if it starts so.

Averaging the likelihood over the current belief gives the evidence
\begin{equation}
  Z(\sigma_e\mid \mathbf{x}, D_t) \;=\; \mathbb{E}_{Q_t}\bigl[L(\sigma_e\mid \mu, \mathbf{w}, \mathbf{x})\bigr]
  \;=\;\sum_{\sigma_T}\mathbb{E}_{\mu}\bigl[P(\sigma_e\mid\sigma_T,\mathbf{x},\mu)\bigr]\,
  \mathbb{E}_{\mathbf{w}}\bigl[\Phi(\sigma_T\, \mathbf{w}\cdot\mathbf{x})\bigr],
\end{equation}
and the Gaussian integrals collapse onto the two scaled fields
\eqref{eq:fields}, giving \eqref{eq:evidence}: the evidence depends on the
receiver's state only through $(\hw,\hmu)$. This is the symmetry that makes the
two sectors mirror images.

The exact Bayesian posterior is not Gaussian. Replacing it by the Gaussian of
maximum entropy relative to the prior, subject to matching the first two moments
of the posterior, gives updates in terms of derivatives of the log evidence,
\begin{equation}
  \mathbf{w}_{t+1} = \mathbf{w}_t + C_t\frac{\partial \ln Z}{\partial \mathbf{w}},
  \qquad
  C_{t+1} = C_t + C_t \frac{\partial^{2}\ln Z}{\partial\mathbf{w}\,\partial\mathbf{w}^{\!\top}} C_t,
\end{equation}
and identically in the trust sector with $\mu_{e|r}, V_{e|r}$ in place of
$\mathbf{w}, C$. Expressing the derivatives in the scaled fields yields
\eqref{eq:update-ideo}--\eqref{eq:update-aff} using the modulation functions
defined in Equations~\eqref{eq:w-modulation} and~\eqref{eq:mu-modulation}. The
first is a Hebbian term $\sigma_e\mathbf{x}$ with a
tensorial, self-annealing rate; the rest is the online estimation of the channel.

\section{Opinion and trust updates under in-context learning}
\label{app:llm}

The two experiments use different tasks because they require different
observables. The opinion experiment controls both fields and locates a decision
threshold. The trust experiment elicits reliability before and after a message.
Both use \texttt{gpt-5.6-luna} at low reasoning effort, with fixed model weights.
The opinion experiment uses $10\,323$ calls and the trust experiment uses $3\,916$.

\subsection{Opinion: \texorpdfstring{$\Fw$}{Fw} by nulling}
\label{app:llm-opinion}

\paragraph{Prompt and design.} The prompt below contains every part of the
opinion experiment. Six fictional records establish the model's initial opinion,
the analyst's track record establishes trust, and the analyst's statement provides
the social evidence whose effect we measure. The later records are a probe: we vary
their number until the model reverses its answer. Fictional settings prevent
factual knowledge from entering the task. We generated sixteen archives, screened
them for position bias, and retained eleven. The displayed condition has four of
six initial records favoring Durnel Lock, an analyst who was correct on sixteen of
twenty questions, and three later records favoring Durnel Lock. The response was
restricted to the two entity names. The prompt is reproduced verbatim.

\textit{System message}
\begin{Verbatim}[fontsize=\scriptsize,breaklines=true,breaksymbolleft={}]
You are an analyst working on a body of historical records. You judge each question only from the briefing you are given, and you always commit to the answer the briefing makes more likely, however slightly.
\end{Verbatim}

\textit{User message}
\begin{Verbatim}[fontsize=\scriptsize,breaklines=true,breaksymbolleft={}]
The registry contains dated engineering entries for locks, sluices, and tow routes maintained by the Sornethic League. You are one of its analysts.

Question: which was opened earlier, Durnel Lock or Pavqet Lock?

You have examined 6 canal entries bearing on it:
  Canal entry 1: indicates that Durnel Lock was opened earlier.
  Canal entry 2: indicates that Durnel Lock was opened earlier.
  Canal entry 3: indicates that Pavqet Lock was opened earlier.
  Canal entry 4: indicates that Pavqet Lock was opened earlier.
  Canal entry 5: indicates that Durnel Lock was opened earlier.
  Canal entry 6: indicates that Durnel Lock was opened earlier.

Registrar Veskul works on the same records. Of the last 20 questions of this kind whose answers were later established beyond doubt, Registrar Veskul answered correctly 16 times and incorrectly 4 times.
Registrar Veskul has now stated: "Durnel Lock was opened earlier."

A later consignment has since reached you: 3 further canal entries bearing on this question, of the same standing as any other record in the collection. All 3 indicate that Durnel Lock was opened earlier.

On the whole of the above, which was opened earlier, Durnel Lock or Pavqet Lock?
\end{Verbatim}

Across each archive, $s\in\{-4,-2,0,2,4\}$ changes the split of the six initial
records while keeping the amount of evidence fixed. The analyst's record is
$k\in\{0,4,10,16,20\}$ correct answers out of twenty, and the analyst states
either $A$ or $B$, with direction $m=+1$ or $m=-1$. Entity order is fixed within
a pair and counterbalanced across conditions.

\paragraph{Measurement.} Let $t$ be the signed number of later records: $t>0$
supports $A$, $t<0$ supports $B$, and $t=0$ adds no record. We increase $|t|$
until the answer reverses and then bisect the final interval. At each value we
sample sixteen binary answers and fit
\begin{equation}
  \Pr(A\mid t)=\Phi\!\left[\beta(t-t^\star)\right],
  \label{eq:nulling-probit}
\end{equation}
so the opinion before the later evidence is $\ell=-t^\star$. We estimate the
threshold with the analyst silent ($\ell_0$) and with the statement present
($\ell_1$); the prompts otherwise differ only by that statement. The full design
contains $11\times5\times5\times2=550$ updates.

Let $H_\lambda$ convert evidence units to the probit fields of
\eqref{eq:fields}. Relative to the analyst's statement,
\begin{equation}
  \hw=mH_\lambda(\ell_0),
  \qquad
  \Delta\hw=m\left[H_\lambda(\ell_1)-H_\lambda(\ell_0)\right].
  \label{eq:nulling-map}
\end{equation}
Thus $\hw>0$ means initial agreement and $\Delta\hw>0$ means movement toward the
statement. The scale $\lambda$ and the mapping from track record to $\hmu$ are
calibrated independently of the opinion updates. Four supplied pieces of evidence
are recovered as $3.6$--$4.1$ pieces, and the inferred testimony weights correlate
with the five stated records at $r=0.99$.

Equation~\eqref{eq:update-ideo} predicts
$\Delta\hw\propto\Fw(\hw,\hmu)$ up to an unobserved susceptibility. We fit one
global positive scale $\alpha_w$ across the $550$ updates; neither the sign
agreement nor the rank correlation in Figure~\ref{fig:llm} depends on this scale.
The result concerns one model at one reasoning effort.

\subsection{Trust: \texorpdfstring{$\Fmu$}{Fmu} by elicited reliability}
\label{app:llm-trust}

\paragraph{Prompt and design.} The prompt below shows the after-message condition
of the trust experiment. The model's opinion and whether it holds that opinion
firmly or tentatively control conviction. The history sets prior trust in the
other agent, and the final message provides new evidence about that agent's
reliability. The before-message prompt is identical except that the paragraph
beginning ``They have now said'' is absent. The model returns an integer from 0
to 100. We generated and froze the materials for twenty open, low-valence themes
before running the experiment. The prompt is reproduced verbatim.

\textit{System message}
\begin{Verbatim}[fontsize=\scriptsize,breaklines=true,breaksymbolleft={}]
You are an agent taking part in a discussion with other agents.
\end{Verbatim}

\textit{User message}
\begin{Verbatim}[fontsize=\scriptsize,breaklines=true,breaksymbolleft={}]
Consider whether difficulty is a necessary part of a good education, or mostly an obstacle that good teaching should remove.

You currently have this opinion on that theme:

  "Difficulty is a necessary part of good education when it is purposeful: grappling with challenging ideas builds understanding, resilience, and independence. However, good teaching should remove unnecessary obstacles so that students struggle with the subject itself, not with confusion, poor design, or lack of support."

You hold that opinion firmly.

You have discussed this theme with another agent 2 times before. On each occasion they said:

  1. "Students learn best when information is presented clearly and directly, so struggle should be eliminated rather than treated as educationally valuable."
  2. "Difficulty is valuable when it encourages students to think, question, and solve problems independently, but ineffective instruction should not be mistaken for rigor."

They have now said:

  "Good education should preserve productive struggle while eliminating avoidable frustration; students learn most deeply when challenges are meaningful and support is available."

What is the chance, from 0 to 100, that this agent's judgement on this theme is sound?
\end{Verbatim}

Across themes and the two conviction instructions, each history is indexed by
$k=n_+-n_-$, where $n_+$ and $n_-$ count statements that agree and disagree with
the model:
\begin{center}
\small
\begin{tabular}{c*{9}{r}}
\toprule
$k$     & $-5$ & $-4$ & $-3$ & $-2$ & $-1$ & $0$ & $1$ & $2$ & $3$ \\
$n_+$   & $0$  & $0$  & $0$  & $0$  & $1$  & $1$ & $2$ & $2$ & $3$ \\
$n_-$   & $5$  & $4$  & $3$  & $2$  & $2$  & $1$ & $1$ & $0$ & $0$ \\
\bottomrule
\end{tabular}
\end{center}
The prompt never labels the agent as reliable. After each history, it gives either
a strongly agreeing ($m=+1$) or strongly disagreeing ($m=-1$) message drawn from
a separate pool.

\paragraph{Measurement.} Before the new message, we elicit the probability $r_0$
that the agent's judgment is sound and the probability $c$ that a separate,
explicitly sound agent would agree with the model; the latter is our direct proxy
for conviction. After the message, we elicit the original agent's reliability
again as $r_m$. We then compute
\begin{equation}
  \hmu=-\Phi^{-1}(r_0),
  \qquad
  \hw=m\Phi^{-1}(c),
  \qquad
  \Delta\hmu=-\Phi^{-1}(r_m)-\hmu.
  \label{eq:trust-readout}
\end{equation}
These are the vertical coordinate, horizontal coordinate, and color in
Figure~\ref{fig:llm}. Each prior context produces two points, one for each $m$.
Four of the $360$ intended contexts lacked the required stimuli, leaving $356$
contexts and $712$ updates.

Equation~\eqref{eq:update-aff} predicts
$\Delta\hmu\propto\Fmu(\hw,\hmu)$ up to an unobserved trust susceptibility. We
fit one global positive scale $\alpha_\mu$ across the $712$ updates; the reported
sign agreement and correlation do not depend on it.

\section{The general group-bias field}
\label{app:cases}

Equation~\eqref{eq:disc-field} is one component of a more general $2\times2$ field $D$ indexed by the classes of receiver and emitter,
$\hw^{D}=\hw+D_{\kappa_r,\kappa_e}$. Since $\kappa_r$ and $\kappa_e$ each take
two values, any such field decomposes uniquely as
\begin{equation}
  D_{\kappa_r\kappa_e} \;=\; a \;+\; b\,\kappa_r\kappa_e
                            \;+\; c\,\kappa_r \;+\; d\,\kappa_e ,
  \label{eq:field-basis}
\end{equation}
Here $a$ is a uniform shift, $b$ depends on whether the two classes match, $c$
depends on the receiver's class, and $d$ depends on the emitter's class. The main
text studies only $b$, with $a=c=d=0$, because it is the only class-dependent
component that is unchanged when the class labels are exchanged.

If each component is carried by a fraction $f_a$, $f_b$, $f_c$, or $f_d$ of the
population, the collective state depends on eight control variables:
$(a,b,c,d,f_a,f_b,f_c,f_d)$. The complete phase diagram is therefore much richer
than the two-dimensional $(b,f_b)$ slice studied in the main text. That slice is
especially informative because it contains four qualitatively distinct
collective phases, so we study it in detail.

\begin{figure}[ht]
  \centering
  \includegraphics[width=0.86\linewidth]{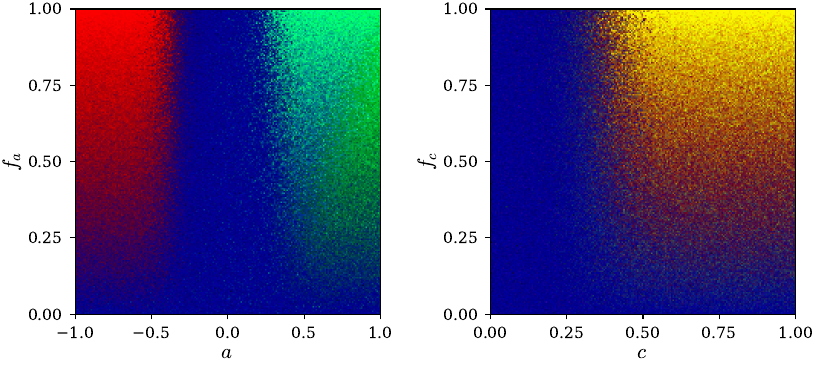}
  \caption{\textbf{Two slices of the general field.} Left: the $(a,f_a)$ plane.
  Right: the $(c,f_c)$ plane. Colors encode the order parameters defined in the
  text.}
  \label{fig:component-phases}
\end{figure}

Figure~\ref{fig:component-phases}
shows two other slices, with all unshown components set to zero. We simulated
these diagrams with $N=40$ on $200\times200$ grids, using one realization per
point. Each composite combines order parameters appropriate to its field. For the
uniform field, they measure population-wide trust or distrust and opinion--trust
polarization. For the receiver-class field, they measure whether credulity follows
class, whether the two classes differ in their internal trust organization, and
opinion--trust polarization.

The uniform field produces widespread distrust for $a<0$, ordinary polarization
near $a=0$ or small $f_a$, and consensus for $a>0$, where widespread trust drives
opinions to align. The receiver-class
field instead drives a transition from ordinary polarization to a state in which
credulity follows class as $c$ and $f_c$ increase. Thus even these two slices have
collective states that do not appear in the $(b,f_b)$ plane.

\section{An unbiased population polarizes into two camps}
\label{app:balance}
\label{app:balance-def}
\label{app:balance-d0}

This appendix supports the polarization analysis in Section~\ref{sec:no-disc}
by using the distributions of pairwise opinion overlap and trust to show directly
that an unbiased population separates into two opposing camps.

\begin{figure}[ht]
  \centering
  \includegraphics[width=1.0\linewidth]{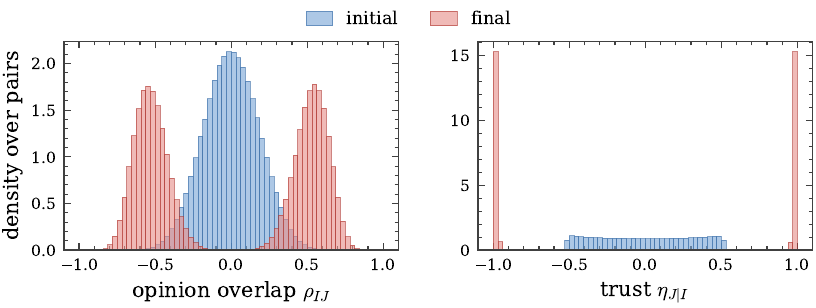}
\caption{\textbf{An unbiased society splits in two.} Eight realizations with
  independently sampled initial conditions, each with $N=400$ agents at $b=0$
  and a simple agenda, before and after $500$
  interactions per ordered pair, with their pairs pooled: $638\,400$ unordered pairs
  for the opinion overlap and $1\,276\,800$ ordered pairs for the trust. Blue is the
  initial state, red the final one. Opinion overlap starts unimodal at zero, of width $1/\sqrt{K}$ as
  random vectors in $K=30$ dimensions require, and ends bimodal; trust starts spread
  across the interior and ends saturated at $\pm1$, so the split in trust is built by
  the dynamics and not supplied by the initial condition. The positive and negative
  peaks correspond to pairs within and across the two camps, respectively.}
  \label{fig:polarization}
\end{figure}

Figure~\ref{fig:polarization} follows the pairwise distributions defined in
Equation~\eqref{eq:pairwise} through an unbiased run. Opinion overlap begins
unimodal around zero and ends with one positive and one negative peak. Trust
likewise develops peaks at $+1$ and $-1$. Positive pairs are aligned and mutually
trusting, while negative pairs hold opposing opinions and distrust one another.
Together, these two populations of pairs show that the agents have separated into
two internally coherent, mutually distrustful camps. Because class labels do not
enter the $b=0$ dynamics, the split is produced by interaction rather than by
class.

\section{Simulation parameters}
\label{app:parameters}

All numerical society experiments use the shared parameters below. Experiment-specific
population sizes and sampling designs are collected in the second table; the same
values are also stated with the corresponding results in the main text and appendices.

\begin{table}[H]
\centering\small
\begin{tabular}{llll}
\toprule
symbol & meaning & value & provenance \\
\midrule
$K$ & dimension of the issue embedding & $30$ & fixed by the agenda-complexity setup \\
$\Delta t$ & interactions per ordered pair & $500$ & stability check (see below) \\
$C_0, V_0$ & initial uncertainties & $I$, $1$ & uninformative prior \\
$\mu_0$ & initial distrust & $\mathcal{U}(-1,1)$ & unstructured initial trust \\
\bottomrule
\end{tabular}
\end{table}

\begin{table}[H]
\centering\scriptsize
\begin{tabular}{@{}lccccc@{}}
\toprule
experiment & $N$ & $P$ & parameter sampling & realizations & updates per realization \\
\midrule
main phase diagram & $100$ & $5$ & $300\times300$ grid in $(b,f_b)$ & $2$/point & $4\,950\,000$ \\
complex-agenda diagram & $40$ & $100$ & $200\times200$ grid in $(b,f_b)$ & $1$/point & $780\,000$\\
general-field slices & $40$ & $5$ & two $200\times200$ grids & $1$/point & $780\,000$ \\
unbiased polarization & $400$ & $5$ & $b=0$ & $8$ total & $79\,800\,000$ \\
finite-size check & $20,30,40,60$ & $5$ & five selected points & $12$/point & $500N(N-1)$ \\
IIIb lifetime test & $40$ & $5$ & selected points & $12$/point & up to $1\,560\,000\,000$ \\
\bottomrule
\end{tabular}
\label{tabela}
\end{table}

In these phase-diagram sweeps, the grid coordinate $f_b$ is the probability
that an individual agent is assigned a group-conditioned bias at initialization.
Biased-agent assignments are sampled independently across agents and realizations.
Consequently, in a finite population the realized number of biased agents satisfies
$n_b\sim\operatorname{Binomial}(N,f_b)$, and the realized fraction
$\hat f_b=n_b/N$ belongs to the discrete set
$\{0,1/N,\ldots,1\}$. Thus, the 300 values of $f_b$ in the main sweep
represent 300 values of the assignment probability, not 300 distinct realizable
population fractions.

At each microscopic update, we sample uniformly with replacement a receiver,
a distinct emitter, and one of the $P$ agenda issues. A run of duration
$\Delta t$ therefore performs
\begin{equation}
  T = \Delta t\,N(N-1)
  \label{eq:simulation-duration}
\end{equation}
updates. Thus $\Delta t=500$ means that each ordered receiver--emitter pair is
selected $500$ times on average. Each ordered pair encounters a particular issue
$500/P$ times on average. The two realizations at each point of the main sweep
therefore aggregate $9\,900\,000$ updates per plotted point.

To reduce Monte Carlo noise between nearby parameter values, societies within a
vectorized batch share the sampled sequence of receiver, emitter, and issue
indices. Their initial opinion weights, distrust variables, agendas, and biased-agent
assignments are sampled separately. We therefore call them realizations rather
than claiming statistically independent interaction schedules.

As agents become more certain, their updates become smaller. We use $\Delta t=500$ because doubling the duration to $1000$
does not change the qualitative collective state or the phase assignment across
four representative points in each phase diagram. The largest absolute percentage change in the order parameters is $3.4\%$ when doubling from $\Delta t=500$ to $\Delta t =1000$.

For the simple agenda shown in the main text, we repeat five
representative points of the plane at $N\in\{20,30,40,60\}$ with twelve
realizations each. Increasing the population from $N=40$ to $N=60$ changes
$\Rmuc$ and $\Rcw$ by at most $0.02$ in absolute value, with no change in phase
assignment. The maps are therefore converged in $N$ at $N=40$, and the cost of
going further is steep. At fixed observation duration $\Delta t$, the number of
updates grows as $N(N-1)$ and hence as $N^2$. If the relaxation duration measured
in interactions per ordered pair grows linearly with $N$, the corresponding cost
of reaching that $N$-dependent timescale would instead grow as $N^3$; the reported
simulations keep $\Delta t=500$ fixed.

\section{Phase diagram for a complex agenda}
\label{app:large-agenda}

\begin{figure}[ht]
  \centering
  \includegraphics[width=1.0\linewidth]{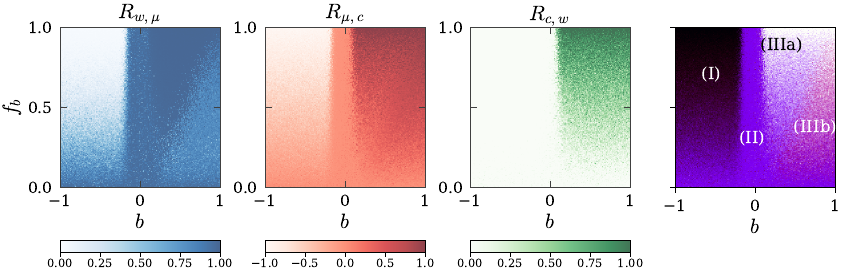}
  \caption{\textbf{Correlations and phase diagram for the complex agenda}
  ($\alpha=3.3$), simulated with $N=40$ on a $200\times200$ grid with one
  realization per point. The first three panels show $\Rwmu$, $\Rmuc$, and $\Rcw$ over
  the $(b,\fb)$ plane; the right panel combines them as RGB channels to identify
  the four collective states. Compared with Figure~\ref{fig:correlations}, $\Rcw$
  now tracks $\Rmuc$ over most of the $b>0$ half, so region~(IIIa) occupies most
  of the discriminatory regime.}
  \label{fig:correlations-large}
  \label{fig:large-agenda}
\end{figure}

This appendix tests whether the phase structure identified in
Section~\ref{sec:results} depends on agenda complexity. We repeat the $(b,\fb)$
sweep for a complex agenda, $\alpha=3.3$, and measure the same three correlations
used for the simple agenda, $\alpha=0.17$. This appendix sweep uses $N=40$ on a
$200\times200$ grid, with one realization and $780\,000$ updates at each point.
Figure~\ref{fig:correlations-large}
shows these correlations and the phase diagram obtained by combining them.

The four states remain in the same arrangement, with the same dependence on
$\fb$ and the same sign change in $b$. The differences are quantitative. Opinion
is no longer capped, so $\Rcw$ rises to meet $\Rmuc$ instead of lagging it, and
region~(IIIa) occupies most of the discriminatory half rather than a narrow band
at its top.

\section{The agenda's orthogonal complement is conserved}
\label{app:conserved}

\begin{figure}
    \centering
    \includegraphics[width=0.5\linewidth]{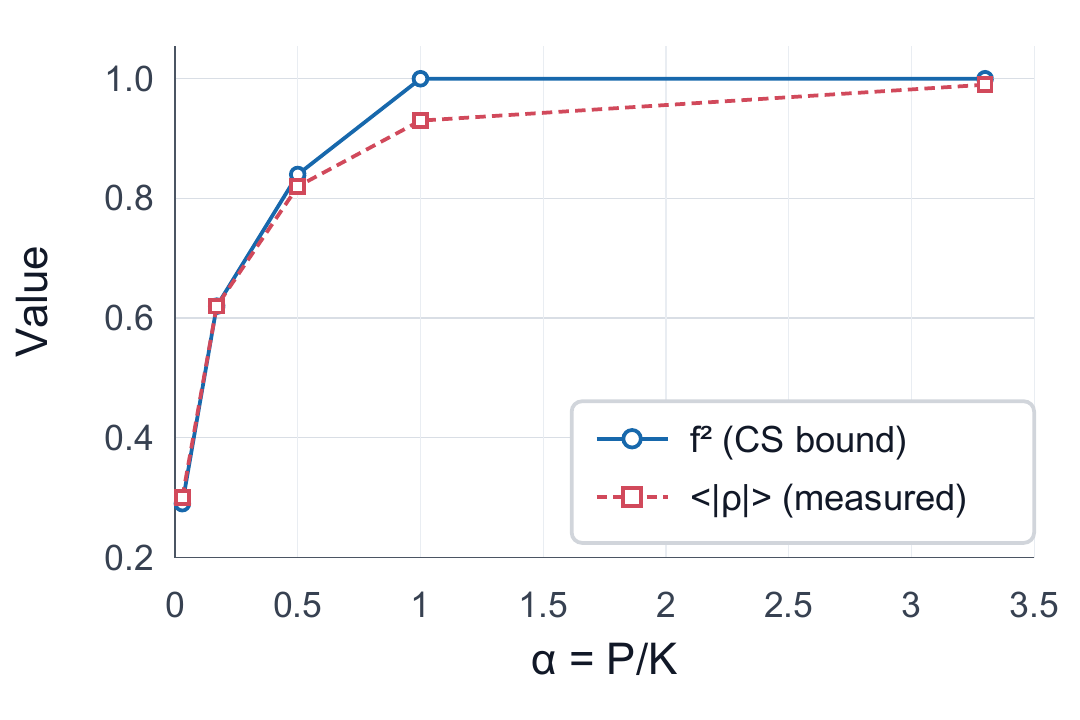}
    \caption{$\langle|\rho|\rangle$ saturates the Cauchy--Schwarz bound of equation \ref{eq:dilution}}
    \label{fig:cauchyschwarz}
\end{figure}

Let $V_P=\mathrm{span}\{\mathbf{x}_1,\dots,\mathbf{x}_P\}$ be the subspace spanned by the
agenda. Equation~\eqref{eq:update-ideo} changes $\mathbf{w}_I$ only within $V_P$, so its
orthogonal component is conserved for every $b$ and $\fb$. Consequently, when
$\alpha=P/K<1$, learning can align only the component $\mathbf{u}_I\in V_P$; the frozen
component $\mathbf{n}_I\perp V_P$ limits the attainable opinion overlap. Writing
$\mathbf{w}_I=\mathbf{u}_I+\mathbf{n}_I$ and the Cauchy--Schwarz inequality gives a bound
\begin{equation}
  |\rho_{IJ}| \;\le\; \, f_I f_J,
  \qquad f_I = \frac{\|\mathbf{u}_I\|}{\|\mathbf{w}_I\|},
  \label{eq:dilution}
\end{equation}
where $f_I$ is the fraction of the weight norm accessible to learning. The resulting
ceiling is saturated in the simulations, figure \ref{fig:cauchyschwarz}.
For $\alpha<1$ the prediction holds to $0.02$. For $\alpha\ge1$ the agenda spans
$\mathbb{R}^{K}$, the complement is empty, and the ceiling disappears, the residual
shortfall being the finite number of interactions rather than a conserved quantity.

\end{document}